\documentstyle[12pt,epsf]{article}

\begin{document}

\font\stepone=cmr10 scaled\magstep1
\font\steptwo=cmr10 scaled\magstep2
\def\s{\smallskip}
\def\ss{\smallskip\smallskip}
\def\b{\bigskip}
\def\bb{\bigskip\bigskip}
\def\bbb{\bigskip\bigskip\bigskip}
\def\i{\item}
\def\ii{\item\item}
\def\iii{\item\item\item}
\def\sqr#1#2{{\vcenter{\vbox{\hrule height.#2pt
 \hbox{\vrule width.#2pt height#1pt \kern#1pt
 \vrule width.#2pt} \hrule height.#2pt}}}}
\def\square{\mathchoice \sqr65 \sqr65 \sqr{2.1}3 \sqr{1.5}3}
\def\utilde{{\tilde u}}
\def\operp{\hbox{${\kern+.25em{\bigcirc}
\kern-.85em\bot\kern+.85em\kern-.25em}$}}
\def\lsim{\;\raise0.3ex\hbox{$<$\kern-0.75em\raise-1.1ex\hbox{$\sim$}}\;}
\def\gsim{\;\raise0.3ex\hbox{$>$\kern-0.75em\raise-1.1ex\hbox{$\sim$}}\;}
\def\no{\noindent}
\def\r{\rightline}
\def\ce{\centerline}
\def\ve{\vfill\eject}
\def\rdots{\mathinner{\mkern1mu\raise1pt\vbox{\kern7pt\hbox{.}}\mkern2mu
 \raise4pt\hbox{.}\mkern2mu\raise7pt\hbox{.}\mkern1mu}}

\def\bu{$\bullet$}
\def\BB{$\bar B B$ }
\def\pp{$\bar p p$ }
\def\BOBO{$\bar B \bar O$BO }
\def\W{W$^{\pm}$ }
\def\ee{$e^+ e^-$ }
\def\Z{$Z^{\circ}$ }

\rightline {UWO/95/DAM/07}
\rightline {November 1995}
\bbb
\ce {\bf CATALYSIS OF DYNAMICAL SYMMETRY BREAKING}
\ce {\bf BY A MAGNETIC FIELD}
\b
\ce { V. A. Miransky\footnote{Based on a talk given at Yukawa
      International Seminar '95 "From the Standard Model to
      Grand Unified Theories", Kyoto, Japan, 21-25 Aug 1995. }}
\s
\ce {\it Institute for Theoretical Physics}
\ce {\it 252143 Kiev, Ukraine}
\ce {\it and}
\ce {\it Department of Applied Mathematics}
\ce {\it University of Western Ontario}
\ce {\it London, Ontario N6A 5B7, Canada}

\bbb
{\bf ABSTRACT.} A constant magnetic field in 3+1
and 2+1 dimensions is a strong catalyst of dynamical chiral
symmetry breaking, leading to the generation of a fermion dynamical mass even
at the weakest attractive interaction between fermions.  The essence
of this effect is the dimensional reduction $D\rightarrow D-2$ in the
dynamics of fermion pairing in a magnetic field.  The effect is
illustrated in the Nambu-Jona-Lasinio model and QED.
Possible applications
of this effect and its extension to inhomogeneous field configurations
are discussed.

\b
\ce {\bf 1.  INTRODUCTION}

At present there are only a few firmly established non-perturbative
phenomena in 2+1 and, especially, 3+1 dimensional field theories.
In this talk, I will describe one more such phenomenon:
dynamical chiral symmetry breaking by a magnetic field. The talk
is
based on a series of the recent papers with V.Gusynin and
I.Shovkovy [1-5].

The problem of fermions in a constant magnetic field had been
considered by Schwinger long ago [6].  In that classical work,
while the interaction with the external magnetic field was considered
in all orders in the coupling constant, quantum dynamics was
treated perturbatively.  There is no spontaneous chiral symmetry
breaking in this approximation.  In the papers [1-5], we reconsidered
this problem, treating quantum dynamics non-perturbatively.  It was
shown that in 3+1 and 2+1 dimensions, a constant magnetic field is a
strong catalyst of dynamical chiral symmetry breaking, leading to
generating a fermion mass even at the weakest attractive interaction
between fermions.  I stress that this effect is universal, i.e.
model independent.

The essence of this effect is the dimensional reduction $D\rightarrow D-2$
in the dynamics of fermion pairing in a magnetic field: while at
$D=2+1$ the reduction is $2+1\rightarrow 0+1$, at $D=3+1$ it is
$3+1\rightarrow 1+1$.  The physical reason of this reduction is the
fact that the motion of charged particles is restricted in directions
perpendicular to the magnetic field.

The emphasis in this talk will be on 3+1 dimensional
field theories.  However, it will be instructive to compare the
dynamics in 2+1 and 3+1 dimensions.

As concrete models for the quantum dynamics, we consider the
Nambu-Jona-Lasinio (NJL) model and QED.  We will show that
the dynamics of the lowest Landau level (LLL) plays the crucial
role in catalyzing spontaneous chiral symmetry breaking.  Actually,
we will see that the LLL plays here the role similar to that of the
Fermi surface in the BCS theory of superconductivity [7].

As was shown in Refs. [4, 5], the dimensional
reduction $D\rightarrow D-2$
is reflected in the structure of the equation describing the
Nambu-Goldstone (NG) modes in a magnetic field.  In Euclidean space,
for weakly interacting fermions, it has the form of a two-dimensional
(one-dimensional) Schr\"odinger equation at $D=3+1~(D=2+1)$:
\begin{eqnarray}
\bigl(-\Delta+m^2_{\rm dyn} + V({\bf r})\bigr) \Psi({\bf r})=0~. %\eqno(1)
\end{eqnarray}
\no Here $\Psi({\bf r})$ is expressed through the Bethe-Salpeter (BS)
function of NG bosons,
\begin{eqnarray*}
\Delta={\partial^2\over \partial x_3^2}+{\partial^2\over \partial x^2_4}
\end{eqnarray*}
\no (the magnetic field is in the $+x_3$ direction, $x_4=it$) for
$D=3+1$, and $\Delta={\partial^2\over \partial x_3^2}$, $x_3=it$, for
$D=2+1$.  The attractive potential $V(\bf r)$ is model dependent.
In the NJL model (both at $D=2+1$ and $D=3+1$), $V({\bf r})$ is a
$\delta$-like potential.  In (3+1)-dimensional ladder QED, the potential
$V({\bf r})$ is
\begin{eqnarray}
V({\bf r})={\alpha\over \pi\ell^2} \exp
\biggl({r^2\over 2\ell^2}\biggr) Ei\biggl(-{r^2\over 2\ell^2}\biggr),
\quad
r^2=x_3^2+x_4^2~, %\eqno(2)
\end{eqnarray}
\no where $Ei(x)=-\int^\infty_{-x} dt \exp(-t)/t$ is the integral
exponential function [8], $\alpha={e^2\over 4\pi}$ is the renormalized
coupling constant and $\ell\equiv |eB|^{-1/2}$ is the magnetic length.
It is important that, as we shall show below, an infrared dynamics is
responsible for spontaneous chiral symmetry breaking in QED in a
magnetic field.  Therefore, because the QED coupling is weak
in the infrared region, the treatment of the non-perturbative dynamics
can be reliable in this problem.

Since $-m^2_{\rm dyn}$ plays the role of energy $E$ in this
equation and $V({\bf r})$ is an attractive potential, the problem is
reduced to finding the spectrum of bound states (with
$E=-m^2_{\rm dyn}<0$) of the Schr\"odinger equation with such a
potential.  More precisely, since only the largest possible value
of $m^2_{\rm dyn}$ defines the stable vacuum [9], we need to find
the lowest eigenvalue of $E$.  For this purpose, we can use results
proved in the literature for the one-dimensional $(d=1)$ and two-dimensional
$(d=2)$ Schr\"odinger equation [10].  These results ensure that there is
at least one bound state for an attractive potential for $d=1$ and $d=2$.
The energy of the lowest level $E$ has the form:
\begin{eqnarray}
E(\lambda)=-m^2_{\rm dyn}(\lambda)=-|eB|f(\lambda)~, %\eqno(3)
\end{eqnarray}
\no where $\lambda$ is a coupling constant ($\lambda$ is $\lambda=G$
in the NJL model and $\lambda=\alpha$ in QED).  While for $d=1$,
$f(\lambda)$ is an analytic function of $\lambda$ at $\lambda=0$, for
$d=2$, it is non-analytic at $\lambda=0$.  Actually we found that,
as $G\rightarrow 0$,
\begin{eqnarray}
m^2_{\rm dyn}=|eB|{N_c^2G^2|eB|\over 4\pi^2}~, %\eqno(4)
\end{eqnarray}
\no where $N_c$ is the number of fermion colors, in (2+1)-dimensional
NJL model [1,2], and
\begin{eqnarray}
m^2_{\rm dyn}={|eB|\over \pi}
\exp\biggl(-{4\pi^2(1-g)\over |eB|N_c G}\biggr)~, %\eqno(5)
\end{eqnarray}
\no where $g\equiv N_cG\Lambda^2/(4\pi^2)$, in (3+1)-dimensional
NJL model [3,5].  In (3+1)-dimensional ladder QED, $m_{\rm dyn}$ is
[4,5]
\begin{eqnarray}
m_{\rm dyn} = C\sqrt{|eB|}\exp \biggl[-{\pi\over2}\biggl({\pi\over2\alpha}
\biggr)^{1/2}\biggr]~, %\eqno(6)
\end{eqnarray}
\no where the constant $C$ is of order one and $\alpha$ is the
renormalized coupling constant relating to the scale $\mu=m_{\rm dyn}$.
The important point is that this expression for $m_{\rm dyn}$ is gauge
invariant.

As we will discuss in Sec.6, there may exist interesting
applications of this effect: in planar condensed matter systems,
in cosmology, in the
interpretation of the heavy-ion scattering experiments, and for
understanding of the structure of the QCD vacuum.
We will also discuss an extension of these results to inhomogeneous
field configurations.

\b
\ce {\bf 2. FERMIONS IN A CONSTANT MAGNETIC FIELD}

In this section we will discuss the problem of relativistic fermions
in a magnetic field in 3+1 dimensions and compare it with the same
problem in 2+1 dimensions.  We will show that the roots of the fact that
a magnetic field is a strong catalyst of chiral symmetry breaking are
actually in this dynamics.

The Lagrangian density in the problem of a relativistic fermion in a
constant magnetic field $B$ takes the form
\begin{eqnarray}
{\cal{L}}={1\over 2}\bigl[\bar\psi,~(i\gamma^\mu D_\mu-m)\psi\bigr]~,
\quad
\mu=0,1,2,3, %\eqno(7)
\end{eqnarray}
\no where the covariant derivative is
\begin{eqnarray}
D_\mu=\partial_\mu-ie A_\mu^{\rm ext}~. %\eqno(8)
\end{eqnarray}
\no We will use the symmetric gauge:
\begin{eqnarray}
A_\mu^{\rm ext} = -{1\over 2} \delta_{\mu 1}Bx_2+{1\over 2}
\delta_{\mu 2}Bx_1 %\eqno(10)
\end{eqnarray}
\no The magnetic field is in the $+x_3$
direction.

The energy spectrum of fermions is [11]:
\begin{eqnarray}
E_n(k_3)=\pm \sqrt{m^2+2|eB|n+k_3^2}~, \quad
n=0,1,2,~\ldots %\eqno(11)
\end{eqnarray}
\no (the Landau levels).  Each Landau level is degenerate: at each
value of the momentum $k_3$, the number of states is
\begin{eqnarray*}
dN_0=S_{12}L_3 {|eB|\over 2\pi}
{dk_3\over 2\pi}
\end{eqnarray*}
\no at n=0, and
\begin{eqnarray*}
dN_n=S_{12}L_3 {|eB|\over \pi}{dk_3\over 2\pi}
\end{eqnarray*}
\no at $n\geq 1$ (where $L_3$ is the size in the $x_3$-direction and
$S_{12}$ is the square in the $x_1x_2$-plane).
In the symmetric gauge (9), the degeneracy is
connected with the angular momentum $J_{12}$ in the $x_1x_2$-plane.

As the fermion mass $m$ goes to zero, there is no energy gap between the
vacuum and the lowest Landau level (LLL) with $n=0$.  The density of
the number of states of fermions on the energy surface with $E_0=0$ is
\begin{eqnarray}
\nu_0=V^{-1} {dN_0\over dE_0}\biggl|_{E_0=0}=
S_{12}^{-1} L_3^{-1}{dN_0\over dE_0}\biggl|_{E_0=0}=
{|eB|\over 4\pi^2}~, %\eqno(11)
\end{eqnarray}
\no where $E_0=|k_3|$ and $dN_0=V{|eB|\over 2\pi}{dk_3\over 2\pi}$
(here $V=S_{12}L_3$ is the volume of the system).  We will see that
the dynamics of the LLL plays the crucial role in catalyzing spontaneous
chiral symmetry breaking.  In particular, the density $\nu_0$ plays the
same role here as the density of states on the Fermi surface $\nu_F$
in the theory of superconductivity [7].

The important point is that the dynamics of the LLL is essentially
(1+1)-dim\-en\-sion\-al.  In order to see this, let us consider the fermion
propagator in a magnetic field.  It was calculated by Schwinger [6]
and has the following form in the gauge (9):
\begin{eqnarray}
S(x,y)=\exp\biggl[{ie\over 2}(x-y)^\mu A_\mu^{\rm ext}(x+y)\biggr]
\tilde S(x-y)~, %\eqno(12)
\end{eqnarray}
\no where the Fourier transform of $\tilde S$ is
\begin{eqnarray}
\tilde S(k) &=& \int^\infty_0
ds\exp\biggl[is\bigl(k_0^2-k_3^2-{\bf k^2}_{\perp}
{\tan (eBs)\over eBs} - m^2\bigr)\biggr] \nonumber\\
&&\hspace{-10mm}\cdot[\bigl(k^0\gamma^0-k^3
\gamma^3+m)(1+\gamma^1\gamma^2 \tan(eBs)\bigr)
-{\bf k}_{\perp}{\mbox{\boldmath$\gamma$}}_{\perp}(1+\tan^2(eBs)\bigr)\bigr]
%\eqno(14)
\end{eqnarray}
\no (here $\bf k_{\perp}=(k_1,k_2),~
{\mbox{\boldmath $\gamma$}}_{\perp}=(\gamma_1,\gamma_2)$).
Then by using the identity $i\tan(x)=   \break
1-2\exp(-2ix)/[1+\exp(-2ix)]$ and
the relation [8]
\begin{eqnarray}
(1-z)^{-(\alpha+1)}\exp\biggl({xz\over z-1}\biggr)=\sum^\infty_{n=0}
L^\alpha_n(x)z^n~, %\eqno(14)
\end{eqnarray}
\no where $L_n^\alpha(x)$ are the generalized Laguerre polynomials,
the propagator $\tilde S(k)$ can be decomposed over the Landau poles
[12]:
\begin{eqnarray}
\tilde S(k)=i \exp\biggl(-{{\bf k}_{\perp}^2\over |eB|}\biggr)
\sum^\infty_{n=0}
(-1)^n {D_n(eB,k)\over k_0^2-k_3^2-m^2-2|eB|n} %\eqno(15)
\end{eqnarray}
\no with
\begin{eqnarray}
D_n(eB,k)&=&(k^0\gamma^0-k^3\gamma^3+m)
\biggl[\bigl(1-i\gamma^1\gamma^2 {\rm sign}(eB)\bigr)L_n
\biggl(2{{\bf k}_{\perp}^2\over |eB|}\biggr) \nonumber\\
&&\hspace{-25mm}-\bigl(1+i\gamma^1\gamma^2{\rm sign}(eB)\bigr)L_{n-1}
\biggl(2{{\bf k}_{\perp}^2\over |eB|}\biggr)\biggr]
+4(k^1 \gamma^1+k^2\gamma^2)L^1_{n-1}
\biggl(2{{\bf k}_{\perp}^2\over |eB|}\biggr)\, ,
\end{eqnarray}

\no where $L_n \equiv L_n^0$ and $L_{-1}^\alpha=0$ by definition.
The LLL pole is
\begin{eqnarray}
\tilde S^{(0)}(k)=i~\exp\biggl(-{{\bf k}_{\perp}^2\over |eB|}\biggr)
{k^0\gamma^0-k^3\gamma^3+m\over k_0^2-k_3^2-m^2}
\bigl(1-i\gamma^1\gamma^2 {\rm sign}(eB)\bigr)~. %\eqno(17)
\end{eqnarray}
\no This equation clearly demonstrates the (1+1)-dimensional
character of the LLL dynamics in the infrared region, with
${\bf k}_{\perp}^2 \ll |eB|$.  Since at $m^2,k_0^2,k_3^2,
{\bf k}_{\perp}^2 \ll |eB|$ the LLL pole dominates in the fermion
propagator, one concludes that the dimensional reduction
$3+1\rightarrow 1+1$ takes place for the infrared dynamics
in a strong (with $|eB| \gg m^2$) magnetic field.  It is clear
that such a dimensional reduction reflects the fact that the
motion of charged particles is restricted in directions perpendicular
to the magnetic field.

The LLL dominance can, in particular, be seen in the calculation of
the chiral condensate:
\begin{eqnarray}
\langle 0|\bar\psi\psi|0\rangle &=&
-\lim_{x\to y} {\rm tr}~S(x,y)=-{i\over (2\pi)^4} {\rm tr}~\int
d^4k~\tilde S_E(k) \nonumber\\
&&\hspace{-30mm}= -{4m\over (2\pi)^4}\int~d^4k \int^\infty_{1/\Lambda^2} ds
\exp\biggl[-s\biggl(m^2+k^2_4+k^2_3+{\bf k}^2_{\perp}
{\tanh(eBs)\over eBs}\biggr)\biggr] \nonumber\\
&&\hspace{-30mm}= -|eB| {m\over 4\pi^2} \int^\infty_{1/\Lambda^2}
{ds\over s}
e^{-sm^2} \coth(|eBs|)
\longrightarrow_{_{_{_{\hspace{-7mm}
{{ m\rightarrow}{ 0}}}}}}
-|eB|{m\over 4\pi^2}
\biggl(\ln{\Lambda^2\over m^2}+O(m^0)\biggr) ,
\end{eqnarray}

\no where $\tilde S_E(k)$ is the image of $\tilde S(k)$ in
Euclidean space and $\Lambda$ is an ultraviolet cutoff.
As it is clear from Eqs. (15) and (17), the logarithmic
singularity in the condensate appears due to the LLL dynamics.

The above consideration suggests that there is a universal mechanism
for enhancing the generation of fermion masses by a strong magnetic
field in 3+1 dimensions: the fermion pairing takes place essentially
for fermions at the LLL and this pairing dynamics is (1+1)-dimensional
(and therefore strong) in the infrared region.  This in turn suggests
that in a magnetic field, spontaneous chiral symmetry breaking takes
place even at the weakest attractive interaction between fermions in
3+1 dimensions.  This
effect was indeed established in the NJL model and QED [3-5].

In conclusion, let us compare the dynamics in a magnetic field in 3+1
dimensions with that in 2+1 dimensions [1,2].  In 2+1 dimensions, the
LLL pole for the propagator of
the four-component fermions [13] is [2,5] :
\begin{eqnarray}
\tilde S^{(0)}(k)=i~\exp\biggl(-{{\bf k}^2\over |eB|}\biggr)
{k^0\gamma^0+m\over k_0^2-m^2}
\bigl(1-i\gamma^1\gamma^2 {\rm sign}(eB)\bigr)~. %\eqno(19)
\end{eqnarray}
\no Then, as $m\rightarrow 0$, the condensate
$\langle 0|\bar\psi\psi|0\rangle$ remains non-zero due to the LLL:
\begin{eqnarray}
\langle 0|\bar\psi\psi|0\rangle =-\lim_{m\to 0}
{m\over (2\pi)^3} \int d^3k
{\exp(-{\bf k}^2/|eB|)\over k_3^2+m^2} = -{|eB|\over 2\pi}
%\eqno(27)
\end{eqnarray}
\no (for concreteness, we consider $m\geq 0$).  The appearance of
the condensate in the flavor (chiral) limit, $m\rightarrow 0$, signals
the spontaneous breakdown of the flavor (chiral) symmetry
even for free fermions in a magnetic field at $D=2+1$ [1,2].
As we will discuss in Section 4, this in turn provides the analyticity
of the dynamical mass $m_{\rm dyn}$ as a function of the coupling constant
$G$ at $G=0$ in the (2+1)-dimensional NJL model.

\b
\ce {\bf 3. IS THE DIMENSIONAL REDUCTION 3+1 $\rightarrow$ 1+1 (2+1
$\rightarrow$ 0+1)}
\ce {\bf CONSISTENT WITH SPONTANEOUS CHIRAL}
\ce {\bf SYMMETRY BREAKING?}

In this section we consider the question whether the dimensional
reduction 3+1 $\rightarrow$ 1+1 (2+1 $\rightarrow$ 0+1) in the dynamics
of the fermion pairing in a magnetic field is consistent with spontaneous
chiral symmetry breaking.  This question occurs naturaly since, due to the
Mermin-Wagner-Coleman (MWC) theorem [14], there cannot be spontaneous
breakdown of continuous symmetries at $D=1+1$ and $D=0+1$.  The MWC theorem
is based on the fact that gapless Nambu-Goldstone (NG) bosons cannot exist
in dimensions less than 2+1.  This is in particular reflected in that the
(1+1)-dimensional propagator of would be NG bosons would lead to infrared
divergences in perturbation theory (as indeed happens in the $1/N_c$
expansion in the (1+1)-dimensional Gross-Neveu model with a continuous
symmetry [15]).

However, the MWC theorem is not applicable to the present problem.  The
central point is that the condensate $\langle 0|\bar\psi\psi|0\rangle$
and the NG modes are {\bf neutral} in this problem and the dimensional
reduction in a magnetic field does not affect the dynamics of the center
of mass of {\bf neutral} excitations.  Indeed, the dimensional reduction
$D\rightarrow D-2$ in the fermion propagator, in the infrared region,
reflects the fact that the motion of {\bf charged} particles is
restricted in the directions perpendicular to the magnetic field.  Since
there is no such restriction for the motion of the center of mass of
neutral excitations, their propagators have $D$-dimensional form in the
infrared region (since the structure of excitations is irrelevant at long
distances, this is correct both for elementary and composite neutral
excitations\footnote{The
Lorentz invariance is broken by a
magnetic field in this problem.  By the $D$-dimensional form, we
understand that the denominator of the propagator depends on energy
and all the components of momentum.  That is, for $D=3+1,~{\cal{D}}(P)\sim
(P_0^2-C_{\perp} {\bf P}_{\perp}^2-C_3P_3^2)^{-1}$ with $C_{\perp},C_3
\not= 0$.}).
This fact was shown for neutral bound states in the NJL model
in a magnetic field, in the $1/N_c$ expansion, in Refs. [2,5].
Since, besides that, the propagator of {\bf massive} fermions is, though
($D-2$)-dimensional, nonsingular at small momenta, the infrared dynamics is
soft in a magnetic field, and spontaneous chiral symmetry breaking is not
washed out by the interactions, as happens, for example, in the
(1+1)-dimensional Gross-Neveu model.

This point is intimately connected with the status of the space-translation
symmetry in a constant magnetic field.  In the symmetric gauge (9),
the translation symmetry along the $x_1$ and $x_2$
directions is broken.  However, for neutral states, all the components
of the momentum of their center of mass are conserved quantum numbers
(this property is gauge invariant).  In order to show
this in the symmetric gauge, let us
introduce the following operators (generators of the group of magnetic
translations) describing the translations in first quantized theory:
\begin{eqnarray}
\hat P_{x_1}={1\over i}{\partial \over \partial x_1}-
{\hat Q\over 2} Bx_2~, \quad
\hat P_{x_2} = {1\over i}{\partial\over \partial x_2}+
{\hat Q\over 2} Bx_1~, \quad
\hat P_{x_3}={1\over i}{\partial\over \partial x_3}  %\eqno (21)
\end{eqnarray}
($\hat Q$ is the charge operator).  One can easily
check that these operators commute with the Hamiltonian of the Dirac
equation in a constant magnetic field.  Also, we get:
\begin{eqnarray}
\bigl[\hat P_{x_1},\hat P_{x_2}\bigr] = {1\over i}
\hat QB~, \quad
\bigl[\hat P_{x_1},\hat P_{x_3}\bigr] =
\bigl[\hat P_{x_2},\hat P_{x_3}\bigr] = 0~. %\eqno(22)
\end{eqnarray}
\no Therefore all the commutators equal zero for neutral states, and
the momentum ${\bf P}=(P_1,P_2,P_3)$ can be used to describe the
dynamics of the center of mass of neutral states.

\b

\ce {\bf 4. THE NAMBU-JONA-LASINIO MODEL IN A MAGNETIC FIELD}

Let us consider the (3+1)-dimensional NJL model with the $U_{\rm L}(1)\times
U_{\rm R}(1)$ chiral symmetry:
\begin{eqnarray}
{\cal{L}}={1\over 2}\bigl[\bar\psi,(i\gamma^\mu D_\mu) \psi\bigr]+
{G\over 2}\bigl[(\bar\psi\psi)^2+
(\bar\psi i\gamma^5\psi)^2\bigr]~, %\eqno(23)
\end{eqnarray}
\no where $D_\mu$ is the covariant derivative (8) and fermion fields
carry an additional ``color" index $\alpha=1,2,\ldots,N_c$.  The theory
is equivalent to the theory with the Lagrangian density
\begin{eqnarray}
{\cal{L}}={1\over 2}\bigl[\bar\psi,(i\gamma^\mu D_\mu)\psi\bigr]-
\bar\psi(\sigma+i\gamma^5\pi)\psi-{1\over 2G}
(\sigma^2+\pi^2)~. %\eqno(24)
\end{eqnarray}
\no The Euler-Lagrange equations for the auxiliary fields $\sigma$
and $\pi$ take the form of constraints:
\begin{eqnarray}
\sigma=-G(\bar\psi\psi)~, \quad
\pi=-G(\bar\psi i\gamma^5\psi)~, %\eqno(25)
\end{eqnarray}
\no and the Lagrangian density (24) reproduces Eq. (23) upon application
of the constraints (25).

The effective action for the composite fields is expressed through the
path integral over fermions:
\begin{eqnarray}
\Gamma(\sigma,\pi)=\tilde\Gamma(\sigma,\pi)-{1\over 2G} \int
d^4x (\sigma^2+\pi^2)~, %\eqno(26)
\end{eqnarray}
\begin{eqnarray}
\exp(i\tilde\Gamma)&=&\int [d\psi][d\bar\psi]
\exp
\left\{
{i\over 2}\int d^4x
\bigl[
\bar\psi,\{i\gamma^\mu D_\mu-(\sigma+i\gamma^5\pi)\}\psi
\bigr]
\right\}  \nonumber\\
&=&
\exp
\left\{
{\rm Tr}\,{\rm Ln}
\biggl[
i\gamma^\mu D_\mu-\bigl( \sigma+i\gamma^5\pi \bigr)
\biggr]
\right\},
\end{eqnarray}
%\no
i.e.
\begin{eqnarray}
\tilde\Gamma(\sigma,\pi)=-i{\rm Tr}\,{\rm Ln}\bigl[i\gamma^\mu D_\mu-
(\sigma+i\gamma^5\pi)\bigr]~. %\eqno(28)
\end{eqnarray}
\no As $N_c\rightarrow \infty$, the path integral over the composite
fields $\sigma$ and $\pi$ is dominated by the stationary points of
the action: $\delta\Gamma/\delta\sigma=\delta\Gamma/\delta\pi=0$.
The dynamics in this limit was analysed [3,5] by using the expansion of
the action $\Gamma$ in powers of derivatives of the composite fields.

Is the $1/N_c$ expansion reliable in this problem?  The answer to this
question is ``yes".  It is connected with the fact, already discussed
in the previous section, that the dimensional reduction 3+1 $\rightarrow$
1+1 by a magnetic field does not affect the dynamics of the center of
mass of the NG bosons.  If the reduction affected it, the $1/N_c$
perturbative expansion would be unreliable.  In particular, the
contribution of the NG modes in the gap equation, in next-to-leading order
in $1/N_c$, would lead to infrared divergences (as happens in the
1+1 dimensional Gross-Neveu model with a continuous chiral symmetry
[15]).
This is not the case here.  Actually, as was shown in Ref.[5],
the next-to-leading order in $1/N_c$ yields small corrections to the
whole dynamics at sufficiently large values of $N_c$. The same also
valid in the 2+1 dimensional NJL model [2].

The effective actions in the 2+1 dimensional and 3+1 dimensional NJL
models in a magnetic field were derived in Refs.[1,2] and Refs.[3,5],
respectively. Here we will discuss the physics underlying the
expressions (4) and (5) for the dynamical mass,for weakly interacting
fermions, in these models.

It is instructive to compare the relation (5) with the relations for
the dynamical mass (energy gap) in the
 (1+1)-dimensional Gross-Neveu model and
in the BCS theory of superconductivity [7].

The relation for $m^2_{\rm dyn}$ in the Gross-Neveu model is
\begin{eqnarray}
m^2_{\rm dyn} = \Lambda^2\exp
\biggl(-{2\pi\over N_cG^{(0)}}\biggr) %\eqno(29)
\end{eqnarray}
\no where $G^{(0)}$ is the bare coupling, which is dimensionless at
$D=1+1$. The similarity between relations (5) and (29) is evident:
$|eB|$ and $|eB|G$ in Eq. (5) play the role of an ultraviolet cutoff
and the dimensionless coupling constant in Eq. (29), respectively.
This of course reflects the point that the dynamics of the fermion
pairing in the (3+1)-dimensional NJL model in a magnetic field is
essentially (1+1)-dimensional.

We recall that, because of the Fermi surface, the dynamics of the
electron in superconductivity is also (1+1)-dimensional.  This analogy
is rather deep.  In particular, the expression (5) for $m_{\rm dyn}$
can be rewritten in a form similar to that for the energy gap
$\Delta$ in the
BCS theory: while
$\Delta \sim \omega_D \exp\bigl(-{\rm const.}/\nu_FG_S\bigr)$, where
$\omega_D$ is the Debye frequency, $G_S$ is a coupling constant and
$\nu_F$ is the density of states on the Fermi surface, the mass
$m_{\rm dyn}$ is $m_{\rm dyn}\sim \sqrt{|eB|}~\exp\bigl(-1/2G\nu_0\bigr)$,
where the density of states $\nu_0$ on the energy surface $E=0$ of
the LLL is now given by expression (11) multiplied by the factor $N_c$.
Thus the energy surface $E=0$ plays here the role of the Fermi surface.

Let us now compare the relation (5) with the relation (4)
in the (2+1)-dimensional
NJL model in a magnetic field, in a weak coupling regime.
While the expression (5) for $m^2_{\rm dyn}$ has an essential
singularity at $G=0$, $m^2_{\rm dyn}$ in the (2+1)-dimensional NJL
model is analytic at $G=0$.  The latter is connected with the fact
that in 2+1 dimensions the condensate $\langle 0|\bar\psi\psi|0\rangle$
is non-zero even for free fermions in a magnetic field (see Eq. (20)).
Indeed, Eq. (25) implies that $m_{\rm dyn}=\langle 0|\sigma|0\rangle=
-G\langle0|\bar\psi\psi|0\rangle$.  From this fact, and Eq. (20),
we get the relation (4), to leading order in $G$.  Therefore the
dynamical mass $m_{\rm dyn}$ is essentially perturbative in $G$ in this
case.

This is in turn connected with the fact that, for $D=2+1$, the
dynamics of fermion pairing in a magnetic field is (0+1)-dimensional.
Indeed,as was already pointed out in Introduction,the dynamics of the
NG modes for $D=2+1$ is described by the one-dimensional $(d=1)$
Schr\"odinger equation (1) with an attractive ($\delta$-like)
potential, where $-m^2_{\rm dyn}$ plays the role of the energy
$E$. The general theorem [10] ensures that, at $d=1$, the energy of
the lowest level $E(G)$ is an analytic function of the coupling
constant $G$ around $G=0$.

On the other hand,the same theorem ensures that the energy $E(G)$
is non-analytic at $G=0$ at $d=2$. Moreover, at $d=2$ for
short-range potentials, the energy $E(G)=-m^2_{\rm dyn}$ takes the
form $E(G)\sim -\exp\bigl[1/(aG)\bigr]$ (with $a$ being
positive constant) as
$G\rightarrow 0$ [10].

Thus the results obtained in the NJL model agree with this general
theorem.

\b
\ce {\bf 5. SPONTANEOUS CHIRAL SYMMETRY BREAKING}
\ce {\bf BY A MAGNETIC FIELD IN QED}

As we indicated in Introduction, in 3+1 dimensional QED, in ladder
approximation, the dynamics of the NG modes is described by the
two-dimensional Schr\"odinger equation (1) with the potential (2).
The essential difference of this
potential with respect to the $\delta$-like potential in
the NJL model is that it is long range.  Indeed, using the
asymptotic relations for $Ei(x)$ [8], we get:
\begin{eqnarray}
V({\bf r}) &\simeq& -{2\alpha\over\pi}{1\over r^2}~, \quad
r\rightarrow \infty~, \nonumber\\
V({\bf r}) &\simeq& -{\alpha\over \pi\ell^2}
\biggl(\gamma+\ln {2\ell^2\over r^2}\biggr)~, \quad
r\rightarrow 0 ~,
\end{eqnarray}
\no where $\gamma\simeq 0.577$ is the Euler constant.  Therefore,
the theorem of Ref.[10] (asserting that, for short-range potentials,
$E(\alpha)\equiv m^2_{\rm dyn}(\alpha) \sim \exp(-1/a\alpha)$, with
$a>0$, as $\alpha\rightarrow 0$) cannot be applied to this case.  As
was shown in Refs.[4,5], the result in this case is:
\begin{eqnarray}
m_{\rm dyn}=C\sqrt{|eB|}\exp\biggl[-{\pi\over2}\biggl(\frac{\pi}{2\alpha}
\biggr)^{1/2}\biggr],%\eqno(31)
\end{eqnarray}
where the constant $C=O(1)$.
Note that this expression is gauge invariant.

Since
\begin{eqnarray}
\lim_{\alpha\rightarrow 0}{\exp\bigl[-1/a\sqrt{\alpha}\bigr]\over
{\exp\bigl[-1/a^\prime \alpha}\bigr]}=\infty~, %\eqno(32)
\end{eqnarray}
\no at $a,a^\prime>0$, we see that the long-range character of the
potential leads to the essential enhancement of the dynamical mass.

The present effect is based on the dynamics of the LLL, i.e., on the
infrared, weakly copling, dynamics in QED. This seems
 suggest that the ladder
approximation is reliable for this problem. However, because of the
(1+1)-dimensional form of the fermion propagator in the infrared
region, there may be also relevant higher order contributions [5].

For example, let us  consider the photon propagator
in a strong ( $|eB|>> m^2_{\rm dyn}, |k^2_{\parallel}|,
|k^2_{\perp}|$ )
magnetic field, with the
polarization operator calculated in one--loop approximation [16].
One can rewrite it in the following form:
\begin{eqnarray}
{\cal
D}_{\mu\nu}&=&-i\Bigg(\frac{1}{k^2}g_{\mu\nu}^{\perp}+
\frac{k_\mu^{\parallel}
k_\nu^{\parallel}}{k^2k^2_{\parallel}}+\nonumber\\
&+&\frac{1}{k^2+k^2_{\parallel}\Pi(k^2_{\parallel})}
(g^{\parallel}_{\mu\nu}-\frac{k^{\parallel}_{\mu}
k^{\parallel}_{\nu}}{k^2_{\parallel}})
-\frac{\lambda}{k^2}\frac{k_\mu k_\nu}{k^2}\Bigg),\\
\Pi(k^2_{\parallel})&=&-\frac{\alpha}{2\pi}\frac{|eB|}{m^2_{\rm dyn}}
\Bigg[\frac{4m^2_{\rm dyn}}{k^2_{\parallel}}
-\frac{8m^4_{\rm dyn}}{k^2_{\parallel}\sqrt{(k^2_{\parallel})^2
-4m^2_{\rm dyn}k^2_{\parallel}}}\times\nonumber\\
&&\times\ln\frac{
\sqrt{
(k^2_{\parallel})^2-4m^2_{\rm dyn}
k^2_{\parallel}}-k^2_{\parallel}}
{\sqrt{(k^2_{\parallel})^2-4m^2_{\rm dyn}
k^2_{\parallel}}+k^2_{\parallel}}\Bigg],
\end{eqnarray}
where the symbols $\perp$ and $\parallel$ are related to the
 $(1,2)$ and
$(0,3)$ components, respectively, and $\lambda$
 is a gauge parameter. The
asymptotic behavior of $\Pi(k^2_{\parallel})$ is:
\begin{eqnarray}
\Pi(k^2_{\parallel})&\to&\frac{\alpha}{3\pi}\frac{|eB|}{m^2_{\rm dyn}}
\qquad \mbox{at}\qquad |k^2_{\parallel}|\ll m^2_{\rm dyn},\\
\Pi(k^2_{\parallel})&\to&-\frac{2\alpha}{\pi}\frac{|eB|}{k^2_{\parallel}}
\qquad \mbox{at}\qquad |k^2_{\parallel}|\gg m^2_{\rm dyn}.
\end{eqnarray}
There is a strong screening effect in the $(g^{\parallel}_{\mu\nu}-
k^{\parallel}_\mu k^{\parallel}_{\nu}/k^2_{\parallel})$--component
of the photon propagator. In particular there is a pole
$-\frac{2\alpha}{\pi}\frac{|eB|}{k^2_{\parallel}}$ in
$\Pi(k^2_{\parallel})$ as $m^2_{\rm dyn}\to 0$: this is of course a
reminiscence of the Higgs effect in the $(1+1)$--dimensional massless
QED (Schwinger model).

Following Refs. [4,5], one can
show that, with the photon propagator (33), the
expression for  $m_{\rm dyn}$ has the form (6) with
$\alpha\to\alpha/2$.

It is a challenge to define the class of all those
diagrams in QED which give a
relevant contribution in this problem.
Since the QED coupling constant is weak in the infrared region, this
seems not to be a hopeless problem\footnote{I thank
V.Gusynin, G.McKeon, T.Sherry, and I.Shovkovy for discussions of this
point.}.

\b
\ce {\bf 6. CONCLUSION}

In 3+1 and 2+1 dimensions, a constant magnetic field is a strong
catalyst of spontaneous chiral symmetry breaking,
leading to the generation of a fermion dynamical mass
even at the weakest attractive interaction between fermions.  The
essence of this effect is the dimensional reduction $D\rightarrow D-2$
in the dynamics of fermion pairing in a magnetic field.

So far we considered the dynamics in the presence of a
constant magnetic field only.  It would be interesting to extend
this analysis to the case of inhomogeneous electromagnetic fields.
In connection with this, note that
in 2+1 dimensions, the present effect is intimately connected with the
fact that the massless Dirac equation in a constant magnetic field admits
an infinite number of normalized solutions with $E=0$ (zero modes)
[1,2].
More precisely, the density of the zero modes
\begin{eqnarray*}
\tilde\nu_0 = \lim_{S\to\infty} S^{-1}N(E)\biggl|_{E=0}
\end{eqnarray*}
\no (where $S$ is a two-dimensional volume of the system) is finite.
As has been already pointed out [2,17],
spontaneous flavor (chiral) symmetry breaking in 2+1
dimensions should be catalysed by all stationary (i.e. independent
of time) field configurations with $\tilde\nu_0$ being finite.  On the
other hand, as we saw in Sec.4 , the density
\begin{eqnarray*}
\nu_0=\lim_{V\to\infty} V^{-1} {dN(E)\over dE}\Biggl|_{E=0}
\end{eqnarray*}
\no of the states with $E=0$ (from a continuous spectrum) plays the crucial
role in the catalysis of chiral symmetry breaking in 3+1 dimensions.
One may expect that the density $\nu_0$ should play an important role
also in the case of (stationary) inhomogeneous configurations in
3+1 dimensions.

As a first step in studying this problem, it would be important to
extend the Schwinger results [6] to inhomogeneous field configurations.
Interesting results in this direction have been recently obtained in
Ref. [18].

In conclusion, let us discuss possible applications of this effect.

Since (2+1)-dimensional relativistic field theories may serve as
effective theories for the description of long wavelength excitations
in planar condensed matter systems [19], this effect may be
relevant for such systems.  It would be also interesting to take
into account this effect in studying the possibility of
the generation of a magnetic field in the vacuum, i.e. spontaneous
breakdown of the Lorentz symmetry, in (2+1)-dimensional QED [20].

In 3+1 dimensions, one potential application of the effect can be
connected with the possibility of the existence of very strong
magnetic fields ($B\sim 10^{24}G$) during the electroweak phase
transition in the early universe [21].  As the results obtained in
this paper suggest, such fields might essentially change the
character of the electroweak phase transition.

Another application of this effect can be connected with the role of
chromomagnetic backgrounds as models for the QCD vacuum (the
Copenhagen vacuum [22]).

Yet another potentially interesting application
is the interpretation of the results of the GSI heavy-ion scattering
experiments in which narrow peaks are seen in the energy spectra of
emitted $e^+e^-$ pairs [23].  One proposed explanation [24] is that
a strong electromagnetic field, created by the heavy ions, induces a
phase transition in QED to a phase with spontaneous chiral symmetry
breaking and the observed peaks are due to the decay of positronium-like
states in the phase.  The catalysis of chiral symmetry breaking by a
magnetic field in QED, studied in this paper, can serve as a toy
example of such a phenomenon.  In order to get a more realistic model,
it would be interesting to extend this analysis to non-constant background
fields [25].

We believe that the effect of the dimensional reduction by external
field configurations may be quite general and relevant for
different non-perturbative phenomena.  It deserves further study.

\b
\ce {\bf ACKNOWLEDGMENTS}

I would like to thank the organizers of Yukawa International Seminar
'95, particularly Taichiro Kugo, for their hospitality and support.

\b
\ce {\bf REFERENCES}
\b
\begin{enumerate}
\item V. P. Gusynin, V. A. Miransky, and I. A. Shovkovy, Phys.
Rev. Lett. {\bf 73} (1994) 3499.
\item V. P. Gusynin, V. A. Miransky, and I. A. Shovkovy, Phys.
Rev. D{\bf 52} (1995) 4718.
\item V. P. Gusynin, V. A. Miransky, and I. A. Shovkovy, Phys.
Lett. B{\bf 349} (1995) 477.
\item V. P. Gusynin, V. A. Miransky, and I. A. Shovkovy, Phys.
Rev. D{\bf 52} (1995) 4747.
\item V. P. Gusynin, V. A. Miransky, and I. A. Shovkovy,
Preprint
UCLA/95/ TEP/26, HEP-PH/9509320.
\item J. Schwinger, Phys. Rev. {\bf 82} (1951) 664.
\item J. Bardeen, L. N. Cooper, and J. R. Schrieffer, Phys. Rev.
{\bf 108} (1957) 1175.
\item I. S. Gradshtein and I. M. Ryzhik, {\it Table of Integrals,
Series and Products} (Academic Press, Orlando, 1980).
\item V. A. Miransky, {\it Dynamical Symmetry Breaking in Quantum
Field Theories} (World Scientific Co., Singapore, 1993).
\item B. Simon, Ann. Phys. {\bf 97} (1976) 279.
\item A. I. Akhiezer and V. B. Berestetsky, {\it Quantum
Electrodynamics} (Interscience, NY, 1965).
\item A. Chodos, I. Everding, and D. A. Owen, Phys. Rev. D{\bf 42}
(1990) 2881.
\item T. Appelquist, M. Bowick, D. Karabali, and L. C. R. Wijewardhana,
Phys. Rev. D{\bf 33} (1986) 3704.
\item N. D. Mermin and H. Wagner, Phys. Rev. Lett. {\bf 17}
(1966) 1133; S. Coleman, Commun. Math. Phys. {\bf 31} (1973) 259.
\item D. Gross and A. Neveu, Phys. Rev. D{\bf 10} (1974)
3235; E. Witten, Nucl. Phys. B{\bf 145} (1978) 110.
\item G. Calucci and R. Ragazzon, J. Phys A: Math. Gen. {\bf 27}
(1994) 2161.
\item R. Parwani, Preprint IP/BBSR/95-12; HEP-TH/9504020.
\item D. Cangemi, E. D'Hoker, and G. Dunne, Phys. Rev. D{\bf 51} (1995)
R2513; Phys. Rev. D{\bf 52} (1995) R3163.
\item R. Jackiw, Phys. Rev. D{\bf 29} (1984) 2375;
I. V. Krive and A. S. Rozhavsky, Sov. Phys. Usp. {\bf 30} (1987) 370;
A. Kovner and B. Rosenstein, Phys. Rev. B{\bf 42} (1990) 4748;
G. W. Semenoff and L. C. R. Wijewardhana, Phys. Rev. D{\bf 45} (1992)
1342; N. Dorey and N. E. Mavromatos, Nucl. Phys. B{\bf 368} (1992) 614.
\item Y. Hosotani, Phys. Lett. B{\bf 319} (1993) 332.
\item T. Vachaspati, Phys. Lett. B{\bf 265} (1991) 258;
K. Enqvist and P. Olesen, Phys. Lett. B{\bf 319} (1993) 178;
J. M. Cornwall, in {\it Unified Symmetry: In the Small and Large},
eds. B. N. Kursunoglu {\it et al.} (Plenum Press, New York, 1995).
\item N. K. Nielsen and P. Olesen, Nucl. Phys. B{\bf 144} (1978)
376.
\item P. Salabura {\it et al.}, Phys. Lett. B{\bf 245} (1990)
153; I. Koeing {\it et al.}, Z. Phys. A{\bf 346} (1993) 153.
\item L. S. Celenza, V. K. Mishra, C. M. Shakin, and
K. F. Lin, Phys. Rev. Lett. {\bf 57} (1986) 55;
D. G. Caldi and A. Chodos, Phys. Rev. D{\bf 36} (1987) 2876;
Y. J. Ng and Y. Kikuchi, Phys. Rev. D{\bf 36} (1987) 2880.
\item D. G. Caldi and S. Vafaeisefat, Phys. Lett. B{\bf 287}
(1992) 185.
\end{enumerate}
\end{document}